\documentclass[a4paper,11pt]{article}
\usepackage{float}
\usepackage{graphicx}
\usepackage{caption, subcaption}
\usepackage{mwe}
\pdfoutput=1 % if your are submitting a pdflatex (i.e. if you have
\usepackage{jinstpub} % for details on the use of the package, please
\usepackage[italian,english]{babel}
\usepackage[T1]{fontenc}
\usepackage[utf8]{inputenc}
\usepackage{graphicx}
\usepackage{wrapfig}
\usepackage{lipsum}  % generates filler text

\title{\boldmath RE3/1 \& RE4/1 RPC chambers integration in the inner region of the forward muon spectrometer in the CMS experiment}

\author[a,1]{E. Voevodina\note{Corresponding author.}}
\author[b]{and I. M. Crotty,}

\affiliation[a]{INFN and Università "Federico II" di Napoli,\\via Cintia 21, Napoli, 80126, Italy}
\affiliation[b]{University of Wisconsin, \\Madison, WI 53706, United States}

\emailAdd{elena.voevodina@cern.ch}

\abstract{The high pseudorapidity ($\eta$) region of the Compact Muon Solenoid (CMS) muon system is covered by Cathode Strip Chambers only and lacks redundant coverage despite the fact that it is a challenging region for muons in terms of backgrounds and momentum resolution.  During the annual Year-End Technical Stops 2022 \& 2023, two new layers of improved Resistive Plate Chambers (iRPC) will be added, RE3/1 \& RE4/1, which will completely cover the region of $1.8 < |\eta| < 2.4$ in the endcap. Thus, the additional new chambers will lead to an increased efficiency for both trigger and offline reconstruction in the difficult region where the background is the highest and the magnetic field is the lowest within the muon system. The extended RPC system will improve the performance and the robustness of the muon trigger. The final design of iRPC chambers and the concept to integrate and install them in the CMS muon system have been finalized. In this report, the main results demonstrating the implementation and installation of the new iRPC detectors in the CMS muon system at high $|\eta|$ region will be presented.}


\arxivnumber{} % only if you have one

\collaboration[c]{on behalf of the CMS Muon Group}

\proceeding{14$^{\text{th}}$ Workshop on Resistive Plate Chambers and Related Detectors\\
 19-23 February 2018\\
  Puerto Vallarta, Jalisco State, MEXICO}

\begin{document}
\maketitle
\flushbottom

\section{Introduction}
\label{sec:intro}

The increase of the energy and luminosity during the future upgrades of the Large Hadron Collider's (LHC) machine will adversely affect the performance of the Compact Muon Solenoid (CMS) muon system due to the harsh background environment and the high pile-up. The CMS Collaboration is currently improving the muon system to maintain the high level of performance achieved during the first and second periods of operation (Run-1 (2010-2012) \&  Run-2 (2015-2018)) also in the challenging environment of the high-luminosity LHC. Operating in the high background particle rate imposes severe restrictions on the gaseous detection technology that can be used: new detector requirements include high rate capability, good spatial resolution for tracking, good time resolution for triggering, and in addition, radiation hardness.  

\begin{wrapfigure}{L}{0.45\textwidth}
\centering
\includegraphics[width=0.45\textwidth]{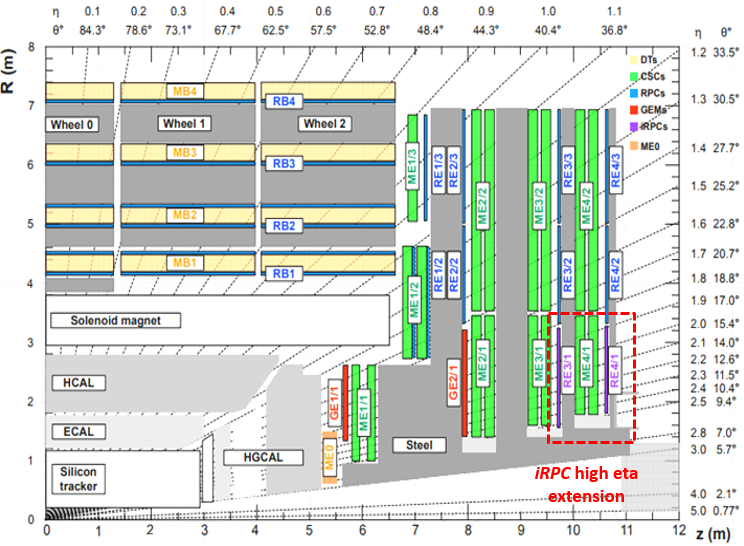}
\caption{A longitudinal quadrant of the CMS detector.}
\vspace{-3.3mm}
\end{wrapfigure}

To fulfil these requirements, new and improved technology of the gaseous detectors have been chosen such as the Gas Electron Multiplier (GEM) and improved Resistive Plate Chambers (iRPC). Therefore, these additional sets of muon detectors have been planned to be install. Before this upgrade, eight RPC layers in total were installed in the four existing endcap disks covering up to the value of pseudorapidity $|\eta|<1.9$. At the moment the very forward region still remains incomplete after first Long Shutdown (LS1) and could be instrumented up to $|\eta| = 2.4$ during LS2 (for GEM project) and annual Year-End Technical Stops 2022 \& 2023 (for iRPC project) (Fig. 1) [1]. The presence of an inner layer in the endcap region equipped with a new improved RPC stations, called RE3/1 \& RE4/1, will increase the overall robustness of the CMS muon spectrometer. The new iRPC RE3/1 and RE4/1 chambers will complement the Cathode Strip Chambers (CSC) stations ME3/1 \& ME4/1 and enhance the local muon measurement by adding track hits. The new hits provided by the new detectors will recover the efficiency losses due to acceptance gaps in this region. The effect will be especially pronounced for high quality muons with hits identified in all four muon stations.

\section{Design of the new improved RPC chambers} 
The 3D-drawing of the new iRPC chambers for inner layers of the endcap stations is shown in Fig. 2 and main geometrical parameters of the RE3/1 and RE4/1 chambers are given in Table 1. To uniformly fill in along the radius and to reduce number of dead areas in the inner layers endcap stations, the design of the new iRPC chambers was chosen identical to the existing wedge-shaped RPC detectors. The new iRPCs will contain a double 1.4 mm high-pressure laminates plate, forming of 1.4 mm thick of the gas gap [2], coated with a conductive graphite paint to form electrodes and insulated from the electrodes by plastic material. In fact, the design will use two identical large gas gaps at the bottom and on top of the chamber with a radially oriented readout panels placed in between. The entire sandwich will be gas tight. The chamber will be contained within a honeycomb box with the chamber services (readout electronics, gas distribution and water cooling circuit) mounted on the outside.

%***************Figures 3 and Table 1 and Table 2 $********************
\begin{figure}[h]
\centering
\begin{minipage}[c]{0.3\textwidth}
\centering\setlength{\captionmargin}{0pt}%
\includegraphics[width=\textwidth]{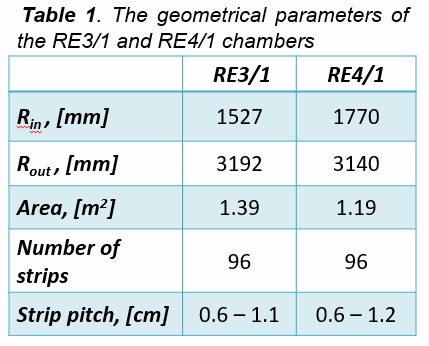}

\end{minipage}%
\hspace{5mm}%
\begin{minipage}[c]{0.3\textwidth}
\centering\setlength{\captionmargin}{0pt}%
\includegraphics[width=\textwidth]{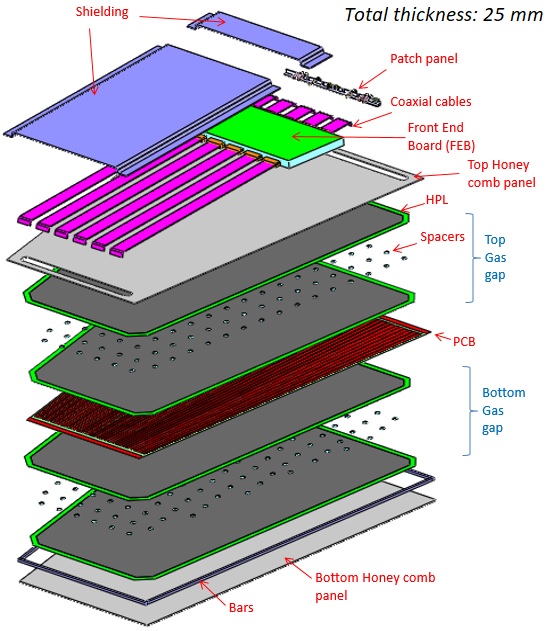}
\caption{ Schematic layout of the iRPC chamber.}
\end{minipage}
\hspace{5mm}%
\begin{minipage}[c]{0.31\textwidth}
\centering\setlength{\captionmargin}{0pt}%
\includegraphics[width=\textwidth]{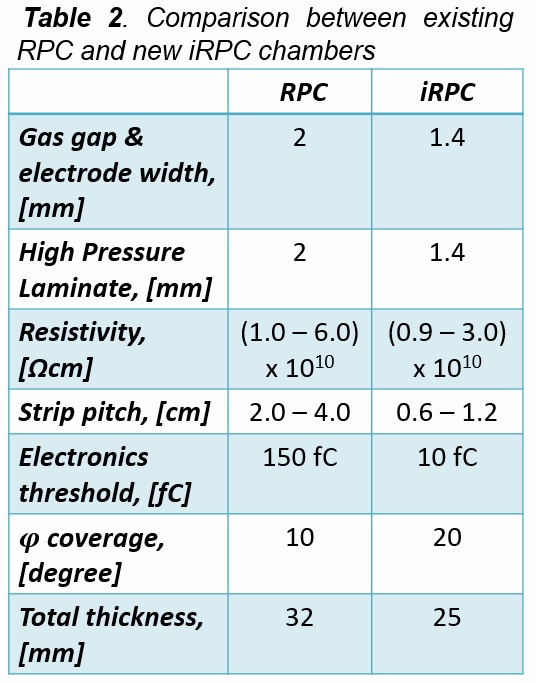}
\end{minipage}%

\vspace{-2mm}
\end{figure}

%**************************************************
The readout panel will consist the signal pick up strips embedded in a readout board made of two parts, i.e. a two large trapezoidal printed circuit boards. Both ends of each strip will be connected through coaxial cables (with the same impedance as the strips) to two different channels of the front-end chip located on the front-end board (FEB). The iRPC's front-end electronics is a 32- channel PETIROC ASIC developed by OMEGA based on the SiGe technology [3], includes broad band fast pre-amplifier and a fast discriminator. Each channel provides a charge measurement and a trigger output that can be used to measure the signal arrival time. The FEB will include the field-programmable gate array device running a Time-to-Digital Converter for the time measurement. The each chamber will divided in 4 $\eta$-partitions, with 24 strips each, yielding a total of 96 strips per chambers. 

The iRPC project will add of 18 new chambers per muon disk, or 72 chambers in total for the RE3/1 and RE4/1 stations in both endcaps [4]. Each station will provide one single hit for muon reconstruction with precise time information of about $2\ ns$, spatial resolution in the transversal direction of about $ 0.3\ cm$ and along strip around $2 \ cm$.

Thus, iRPC technology, as compared with the present RPC one, will have a number of advantages: the thinner electrodes and a narrower gas gap will reduce the recovery time of the electrodes and also the total charge produced in an ionization avalanche; the reduction of the integrated charge deposited will slow down the aging process; the resulting loss in gas gain will be compensated by the higher signal amplification of new front-end electronics; in addition, the operational high voltage will be lower also. All differences between existing RPC and new iRPC chambers are presented in Table 2.

\section{ Installation \& integration of iRPC chambers in the endcap region}

%\begin{wrapfigure}{r}{0.4\textwidth} %this figure will be at the right
%\vspace{-5mm}
%\begin{center}
%\includegraphics[width=0.4\textwidth]{example-image-a}
%\end{center}
%\vspace{-5mm}
%\caption { Schematic view of the RE4/1 chambers mounted on the endcap disk. }
%\vspace{-3mm}
%\end{wrapfigure}

The RE3/1 chambers will be mounted directly on the endcap yoke 3 (YE3) iron disk (Fig. 3), using the foreseen mounting points threaded into the yoke steel. In this case, they will cover the circular neutron shielding attached to the inner part of YE3 and reach the cylindrical neutron shielding surrounding the flange that separates the yokes YE2 and YE3 (Fig. 4-top). 

\begin{figure}[!h]
	\centering
	\begin{minipage}[t]{0.45\linewidth}
		\centering
		\includegraphics[width=\linewidth]{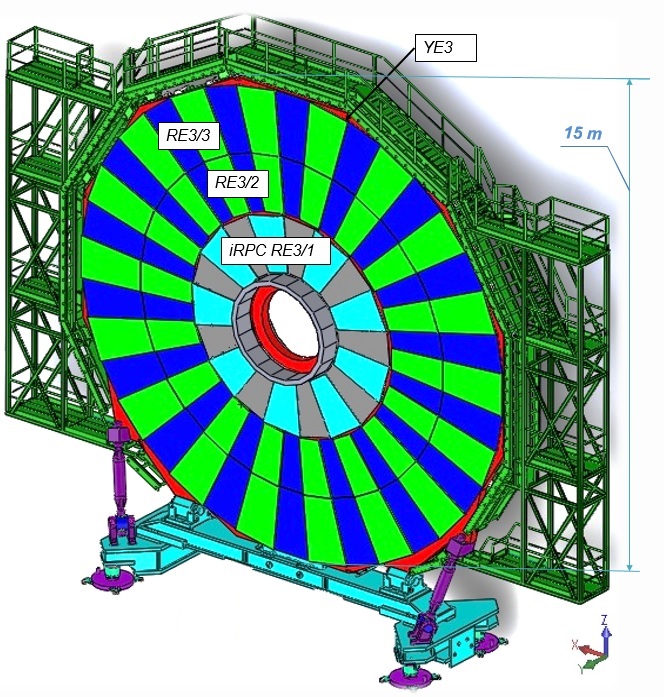}
		\caption{3D-drawing of the RE3/1 chambers fixed on the YE3.}
	\end{minipage}
	\hspace{3mm}
	\begin{minipage}[t]{0.5\linewidth}
		\centering
		\includegraphics[width=\linewidth]{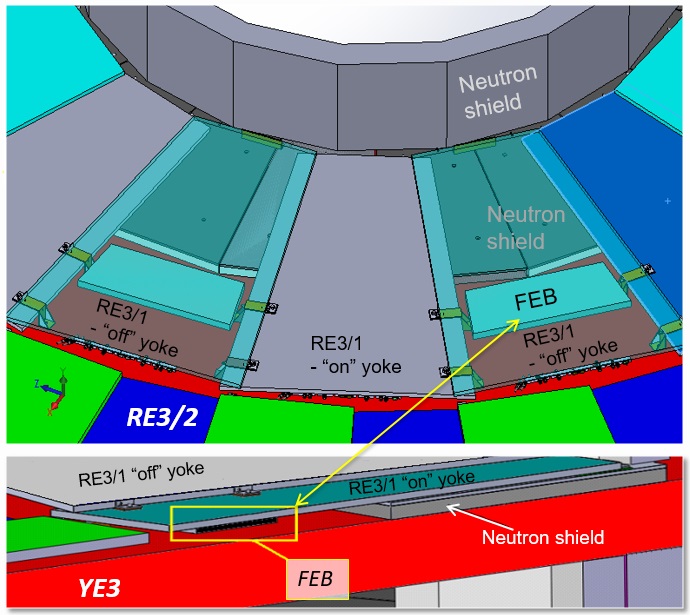}
		\caption{Top: Detailed scheme of installation of the RE3/1 chambers on the YE3. Bottom: The FEBs mounted behind RE3/1 chambers }
	\end{minipage}

\vspace{-2mm}
\end{figure}

Due to the extension of the $|\eta|$ coverage and reduction of the available space between iRPC and CSC chambers, the FEBs will be fixed behind iRPC chambers (Fig. 4-bottom). Access will require the removal of the chamber.

\begin{wrapfigure}{r}{0.45\textwidth} %this figure will be at the right
\vspace{-10mm}
\begin{center}
\includegraphics[width=0.45\textwidth]{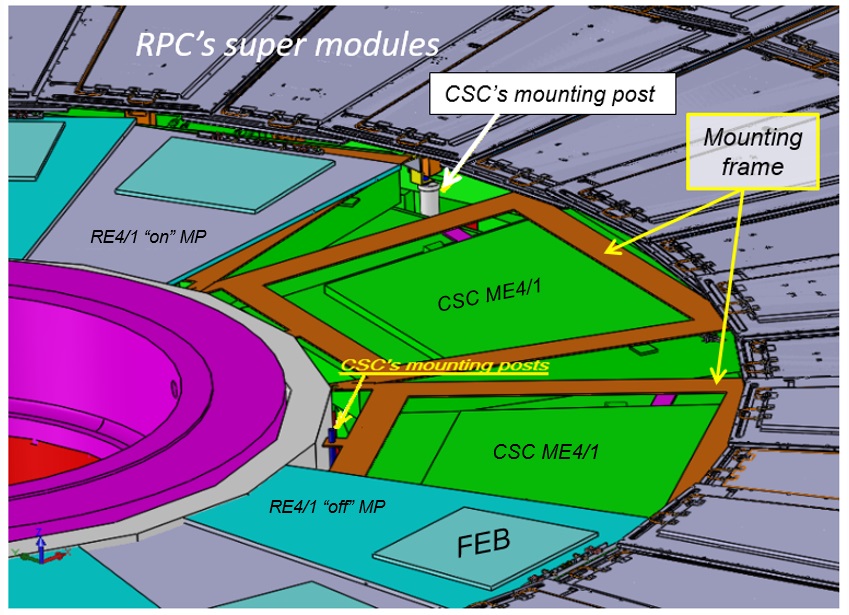}
\end{center}
\vspace{-5mm}
\caption { Schematic view of the RE4/1 chambers mounted on the mounting plate. }
\vspace{-12mm}
\end{wrapfigure}

 % During the installation of the chamber, the $" Z"$ and  $" L"$ - shape mounting brackets ( Fig.3 - bottom) will be used. Due to the lack space between iRPC and CSC chambers (only $20\ mm$ is the available space ) and to decrease the thermal load, coming from CSC's electronics, iRPC's frond-end electronics will be fixed underneath chambers (in the space between iRPC chambers and YE3).

RE4/1 chambers will be installed in a high $|\eta|$ region, close to the RPC's super modules, over the ME4/1 chambers. In this case, aluminum mounting frame will be installed back of the CSCs, to which the iRPCs will be attached (Fig. 5). The chambers will be directly screwed to this frame, iRPCs and frames will be fixed separately.

The RE3/1 and RE4/1 hight voltage (HV) power system will be accommodated in the present RPC's HV racks locatied in the underground service cavern, therefore no new main frames will be required. The RE3/1 and RE4/1 low voltage (LV) power will be supplied by an extension of the present RPC’s LV power system, located mainly in the underground experiment cavern [4].
The new iRPC chambers will use the same $C_{2}H_{2}F_{6}/iC_{4}H_{10}/SF_{6} \ (96.2/3.5/0.3)$ gas mixture as now at present. The current gas distribution system will be extended by adding the new pipings and bulkheads around the yoke.

%All FEBs will be cooled by circulating water from the Endcap cooling circuit. All services (pipe and cable routings) have been planned to instal next year during the LS2.

\section{Conclusions}
The R\&D activity has been performed by the Resistive Plate Chamber (RPC) group of the Compact Muon Solenoid in order to develop an improved RPC (iRPC) that fulfills the CMS requirements for high-luminosity LHC. The baseline of the new 1.4 mm thick iRPC chamber design was chosen to be similar to the present trapezoidal RPC detectors, which are using successfully in the endcap region. The RE3/1 and RE4/1 chambers are planned to be installed during the yearly technical stops at the end of 2022 and 2023. The RE3/1 chambers will be directly fixed to the front face of 3rd endcap yoke in the high pseudorapidity region using the simplified kinematic mounts. The RE4/1 installation will be more involved than the RE3/1. All exisitng services infrastruction will be extended to cover the needs of the iRPC project. The iRPC's services will be installed and mounted during Long Shutdown-2 in 2019-2020.

%The RE3/1 and RE4/1 HV power system will be accommodated in the actual RPC HV rack locating in the underground service cavern, therefore no new main frames will be required. The RE3/1 and RE4/1 LV power system will be an extension of the present RPC LV system, located mainly in the underground experiment cavern.
%All exisitng services infrastruction will be extended to cover the needs of the iRPC project. The iRPC's services will be installed and mounted during Long Sutdown 2 (LS2) in 2019-2020.

\acknowledgments

Many thanks to all our colleagues from the CMS RPC Collaboration for their contributions to the upgrade RPC project.


\end{document}